%Paper: hep-th/9509126
%From: Krzysztof Pilch <pilch@physics1.usc.edu>
%Date: Fri, 22 Sep 1995 15:19:12 -0700 (PDT)

%--------------------------------------------------------------------------
%
%    THE BV-ALGEBRA STRUCTURE OF W_3 COHOMOLOGY
%
%     Peter Bouwknegt and Krzysztof Pilch
%
%
%  published in the proceedings of
%  "Gursey Memorial Conference I: Strings and Symmetries,"
%  Istanbul, June 1994,
%  eds. G. Aktas et al., Lect. Notes in Phys. 447,
%  (Springer Verlag, Berlin, 1995)
%
%  (USC-94/17, September 1994)
%
%  TeX-file   (needs \input tables.tex and amssym.def (version>2.1))
%--------------------------------------------------------------------------
\baselineskip=1.2\baselineskip
\nopagenumbers
\footline{\hss \tenrm -- \folio\ -- \hss}
\hfuzz=10pt
%%%%%%%%%%%%%%%%%%%%%%%%%%%%%%%%%%%%%%%%%%%%%%%%%%%%%%%%%%%%%%%%%%%%%%%%%%%%%%
%%    mutilated harvmac + ams_option
%---------------------------------------------------------------------
\catcode`\@=11 % This allows us to modify PLAIN macros.

%       use \nolabels to get rid of eqn, ref, and fig labels in draft mode
\def\nolabels{\def\wrlabel##1{}\def\eqlabel##1{}\def\reflabel##1{}}
\def\writelabels{\def\wrlabel##1{\leavevmode\vadjust{\rlap{\smash%
{\line{{\escapechar=` \hfill\rlap{\sevenrm\hskip.03in\string##1}}}}}}}%
\def\eqlabel##1{{\escapechar-1\rlap{\sevenrm\hskip.05in\string##1}}}%
\def\thlabel##1{{\escapechar-1\rlap{\sevenrm\hskip.05in\string##1}}}%
\def\reflabel##1{\noexpand\llap{\noexpand\sevenrm\string\string\string##1}}}
\nolabels
%
% tagged sec numbers
\global\newcount\secno \global\secno=0
\global\newcount\meqno \global\meqno=1
\global\newcount\mthno \global\mthno=1
\global\newcount\mexno \global\mexno=1
\global\newcount\mquno \global\mquno=1
\def\newsec#1{\global\advance\secno by1 %\message{(\the\secno. #1)}
\global\subsecno=0\xdef\secsym{\the\secno.}\global\meqno=1\global\mthno=1
\global\mexno=1\global\mquno=1
%\ifx\answ\bigans \vfill\eject \else \bigbreak\bigskip \fi  %if desired
\bigbreak\medskip\noindent{\bf\the\secno. #1}\writetoca{{\secsym} {#1}}
\par\nobreak\medskip\nobreak}
\xdef\secsym{}
\global\newcount\subsecno \global\subsecno=0
\def\subsec#1{\global\advance\subsecno by1%\message{(\secsym\the\subsecno. #1)}
\bigbreak\noindent{\it\secsym\the\subsecno. #1}\writetoca{\string\quad
{\secsym\the\subsecno.} {#1}}\par\nobreak\medskip\nobreak}
\def\appendix#1#2{\global\meqno=1\global\mthno=1\global\mexno=1
\global\subsecno=0
\xdef\secsym{\hbox{#1.}}
\bigbreak\bigskip\noindent{\bf Appendix #1. #2}%\message{(#1. #2)}
\writetoca{Appendix {#1.} {#2}}\par\nobreak\medskip\nobreak}
%
%       \eqn\label{a+b=c}       gives displayed equation, numbered
%                               consecutively within sections.
%     \eqnn and \eqna define labels in advance (of eqalign?)
%
\def\eqnn#1{\xdef #1{(\secsym\the\meqno)}\writedef{#1\leftbracket#1}%
\global\advance\meqno by1\wrlabel#1}
\def\eqna#1{\xdef #1##1{\hbox{$(\secsym\the\meqno##1)$}}
\writedef{#1\numbersign1\leftbracket#1{\numbersign1}}%
\global\advance\meqno by1\wrlabel{#1$\{\}$}}
\def\eqn#1#2{\xdef #1{(\secsym\the\meqno)}\writedef{#1\leftbracket#1}%
\global\advance\meqno by1$$#2\eqno#1\eqlabel#1$$}
%
%          theorems and examples
%
\def\thm#1{\xdef #1{\secsym\the\mthno}\writedef{#1\leftbracket#1}%
\global\advance\mthno by1\wrlabel#1}
\def\exm#1{\xdef #1{\secsym\the\mexno}\writedef{#1\leftbracket#1}%
\global\advance\mexno by1\wrlabel#1}
%
%                        footnotes
\newskip\footskip\footskip14pt plus 1pt minus 1pt %sets footnote baselineskip
\def\f@@t{\baselineskip\footskip\bgroup\aftergroup\@foot\let\next}
\setbox\strutbox=\hbox{\vrule height9.5pt depth4.5pt width0pt}
\global\newcount\ftno \global\ftno=0
\def\foot{\global\advance\ftno by1\footnote{$^{\the\ftno}$}}
%
%say \footend to put footnotes at end
%will cause problems if \ref used inside \foot, instead use \nref before
\newwrite\ftfile
\def\footend{\def\foot{\global\advance\ftno by1\chardef\wfile=\ftfile
$^{\the\ftno}$\ifnum\ftno=1\immediate\openout\ftfile=foots.tmp\fi%
\immediate\write\ftfile{\noexpand\smallskip%
\noexpand\item{f\the\ftno:\ }\pctsign}\findarg}%
\def\footatend{\vfill\eject\immediate\closeout\ftfile{\parindent=20pt
\centerline{\bf Footnotes}\nobreak\bigskip\input foots.tmp }}}
\def\footatend{}
%
%     \ref\label{text}
% generates a number, assigns it to \label, generates an entry.
% To list the refs on a separate page,  \listrefs
%
\global\newcount\refno \global\refno=1
\newwrite\rfile
\def\ref{\the\refno\nref}
\def\bref{\nref}
\def\nref#1{\xdef#1{\the\refno}\writedef{#1\leftbracket#1}%
\ifnum\refno=1\immediate\openout\rfile=refs.tmp\fi
\global\advance\refno by1\chardef\wfile=\rfile\immediate
\write\rfile{\noexpand\item{[#1]\ }\reflabel{#1\hskip.31in}\pctsign}\findarg}
%       horrible hack to sidestep tex \write limitation
\def\findarg#1#{\begingroup\obeylines\newlinechar=`\^^M\pass@rg}
{\obeylines\gdef\pass@rg#1{\writ@line\relax #1^^M\hbox{}^^M}%
\gdef\writ@line#1^^M{\expandafter\toks0\expandafter{\striprel@x #1}%
\edef\next{\the\toks0}\ifx\next\em@rk\let\next=\endgroup\else\ifx\next\empty%
\else\immediate\write\wfile{\the\toks0}\fi\let\next=\writ@line\fi\next\relax}}
\def\striprel@x#1{} \def\em@rk{\hbox{}}

\def\addref#1{\immediate\write\rfile{\noexpand\item{}#1}} %now unnecessary
\def\startrefs#1{\immediate\openout\rfile=refs.tmp\refno=#1}
\def\xref{\expandafter\xr@f}\def\xr@f[#1]{#1}
\def\refs#1{[\r@fs #1{\hbox{}}]}
\def\r@fs#1{\edef\next{#1}\ifx\next\em@rk\def\next{}\else
\ifx\next#1\xref #1\else#1\fi\let\next=\r@fs\fi\next}
%

%
% this is ugly, but moore insists
\newwrite\ffile\global\newcount\figno \global\figno=1
\def\fig{fig.~\the\figno\nfig}
\def\nfig#1{\xdef#1{fig.~\the\figno}%
\writedef{#1\leftbracket fig.\noexpand~\the\figno}%
\ifnum\figno=1\immediate\openout\ffile=figs.tmp\fi\chardef\wfile=\ffile%
\immediate\write\ffile{\noexpand\medskip\noexpand\item{Fig.\ \the\figno. }
\reflabel{#1\hskip.55in}\pctsign}\global\advance\figno by1\findarg}
\def\xfig{\expandafter\xf@g}\def\xf@g fig.\penalty\@M\ {}
\def\figs#1{figs.~\f@gs #1{\hbox{}}}
\def\f@gs#1{\edef\next{#1}\ifx\next\em@rk\def\next{}\else
\ifx\next#1\xfig #1\else#1\fi\let\next=\f@gs\fi\next}
\newwrite\lfile
{\escapechar-1\xdef\pctsign{\string\%}\xdef\leftbracket{\string\{}
\xdef\rightbracket{\string\}}\xdef\numbersign{\string\#}}

\def\writestop{\def\writestoppt{\immediate\write\lfile{\string\pageno%
\the\pageno\string\startrefs\leftbracket\the\refno\rightbracket%
\string\def\string\secsym\leftbracket\secsym\rightbracket%
\string\secno\the\secno\string\meqno\the\meqno}\immediate\closeout\lfile}}
\def\writestoppt{}\def\writedef#1{}
\def\seclab#1{\xdef #1{\the\secno}\writedef{#1\leftbracket#1}\wrlabel{#1=#1}}
\def\subseclab#1{\xdef #1{\secsym\the\subsecno}%
\writedef{#1\leftbracket#1}\wrlabel{#1=#1}}
\newwrite\tfile \def\writetoca#1{}
\def\leaderfill{\leaders\hbox to 1em{\hss.\hss}\hfill}
%       use this to write file with table of contents
\def\writetoc{\immediate\openout\tfile=toc.tmp
   \def\writetoca##1{{\edef\next{\write\tfile{\noindent ##1
   \string\leaderfill {\noexpand\number\pageno} \par}}\next}}}
%       and this lists table of contents on second pass

%
\catcode`\@=12 % at signs are no longer letters
%
%---------------------------------------------------------------------
%%%%%%%%%%%%%%%%%%%%%%%%%%%%%%%%%%%%%%%%%%%%%%%%%%%%%%%%%%%%%%%%%%%%%%%%
%
%    Here our macros start
%
%%%%%%%%%%%%%%%%%%%%%%%%%%%%%%%%%%%%%%%%%%%%%%%%%%%%%%%%%%%%%%%%%%%%%%%%

\def\vev#1{\langle #1 \rangle}

\def\darr#1{\raise1.5ex\hbox{$\leftrightarrow$}\mkern-16.5mu #1}
\def\half{{\textstyle{1\over2}}} %puts a small half in a displayed eqn

%
%  Greek abbreviations (to be used within $....$)
%
\def\al{\alpha}
\def\be{\beta}
  
\def\de{\delta}  \def\De{\Delta}

\def\et{\eta}
\def\th{\theta}  \def\Th{\Theta}

\def\la{\lambda} \def\La{\Lambda}
\def\rh{\rho}
\def\si{\sigma}

\def\ph{\phi}

%
%  boldface abbreviations  (to be used within $..$)
%

%

%
% Calligraphic abbreviations (to be used within $..$)
%
\def\cA{{\cal A}} 
 \def\cD{{\cal D}}
 
\def\cH{{\cal H}}
\def\cI{{\cal I}}
\def\cL{{\cal L}}
\def\cM{{\cal M}}

\def\cP{{\cal P}}
 \def\cS{{\cal S}}

\def\cW{{\cal W}}

\def\mapright#1{\smash{\mathop{\longrightarrow}\limits^{#1}}}
\def\vev#1{\langle #1 \rangle}

\def\ie{{\it i.e.\ }}
\def\eg{{\it e.g.\ }}

%----------------------------------------------------------------------------
% SOME ADDITIONAL MATH SYMBOLS
%
\input amssym.def

\def\CC{{\Bbb C}}
\def\ZZ{{\Bbb Z}}

\def\QQ{{\Bbb Q}}

\def\bfg{{\frak g}}
\def\bfh{{\frak h}}

\def\sltw{\frak{sl}_2}  
\def\slth{\frak{sl}_3}

%
%------------------------------------------------------------------------
% JOURNALS
%

\def\AnM#1{Ann.\ Math.\ {\bf #1}}

\def\CMP#1{Comm.\ Math.\ Phys.\ {\bf #1}}

\def\IJMP#1{Int.\ J.\ Mod.\ Phys.\ {\bf #1}}

\def\LMP#1{Lett.\ Math.\ Phys.\ {\bf #1}}
\def\LNM#1{Lect.\ Notes in Math.\ {\bf #1}}
\def\MPL#1{Mod.\ Phys.\ Lett.\ {\bf #1}}
\def\NPB#1{Nucl.\ Phys.\ {\bf B#1}}
\def\PLB#1{Phys.\ Lett.\ {\bf {#1}B}}

\def\PRep#1{Phys.\ Rep.\ {\bf #1}}

%

%
%%%%%%%%%%%%%%%%%%%%%%%%%%%%%%%%%%%%%%%%%%%%%%%%%%%%%%%%%%%%%%%%%%%%%%%%%%
%%%%
%%%%   END MACRO FILE ISTMAC.TEX
%%%%
%%%%%%%%%%%%%%%%%%%%%%%%%%%%%%%%%%%%%%%%%%%%%%%%%%%%%%%%%%%%%%%%%%%%%%%%%%

\input tables

%\writelabels
%--------------------------------------------------------------------------
%
%   ADDITIONAL DEFINITIONS
%
\def\Wt{{\cal W}_3} 
   
\def\ffM{ \hbox{$M$}\kern-.9em\hbox{$\overline{\phantom{N}}$}}
\def\ffF{ \hbox{$F$}\kern-.53em\hbox{$\overline{\phantom{I}}$}}
\def\ghb#1{b^{[#1]}} \def\ghc#1{c^{[#1]}}
%
%--------------------------------------------------------------------------
%
%   REFERENCES
%

\bref\BMPa{
P.~Bouwknegt, J.~McCarthy and K.~Pilch, \LMP{29} (1993) 91
({\tt hep-th/9302086}).}

\bref\BMPb{
P.~Bouwknegt, J.~McCarthy and K.~Pilch, in ``Perspectives in Mathematical
Physics,'' Vol.\ III, {\it eds.} R.~Penner and S.T.~Yau, pp.\ 77--89,
(International Press, Boston, 1994) ({\tt hep-th/9303164}).}

\bref\BMPc{
P.~Bouwknegt, J.~McCarthy and K.~Pilch, in the proceedings of the workshop
``Strings, Conformal Models and Topological Field Theory,'' Carg\`ese 1993,
{\it eds.\ } L.~Baulieu et al.,
(Plenum Press, New York, 1995) ({\tt hep-th/9311137}).}

\bref\BMPd{
P.~Bouwknegt, J.~McCarthy and K.~Pilch,
{\it The $\cW_3$ algebra: modules, semi-infinite cohomology and
BV-structure}, USC-95/18, ADP-95-46/M38 {\tt (hep-th/9509119)}.}

\bref\BS{
P.~Bouwknegt and K.~Schoutens, \PRep{223} (1993) 183 ({\tt hep-th/9210010}).}

\bref\TM{
J.~Thierry-Mieg, \PLB{197} (1987) 368.}

\bref\BLNW{
M.~Bershadsky, W.~Lerche, D.~Nemeschansky and N.P.~Warner,
\PLB{292} (1992) 35 ({\tt hep-th/9207067});
\NPB{401} (1993) 304 ({\tt hep-th/9211040}).}

\bref\BGG{
I.N.~Bernstein, I.M.~Gel'fand and S.I.~Gel'fand,
in ``Lie groups and their representations,'' Proc.\ Summer School in
Group Representations, Bolyai Janos Math.\ Soc., Budapest 1971, pp.\
21-64, (New York, Halsted, 1975).}

\bref\Zh{
D.P.~Zhelobenko, ``Compact Lie groups and their representations,''
(Providence, Amer.\ Math.\ Soc., 1973).}

\bref\Wi{
E.~Witten, \NPB{373} (1992) 187 ({\tt hep-th/9108004});
E.~Witten and B.~Zwiebach, \NPB{377} (1992) 55 ({\tt hep-th/9201056}).}

\bref\WY{
Y.-S.~Wu and C.-J.~Zhu, \NPB{404} (1993) 245 ({\tt hep-th/9209011}).}

\bref\LZ{
B.H.~Lian and G.J.~Zuckerman, \CMP{154} (1993) 613 ({\tt hep-th/9211072}).}

\bref\Ge{
M.~Gerstenhaber, \AnM{78} (1962) 267.}

\bref\Ko{
J.-L.~Koszul, Ast\'erisque (hors s\'erie) (1985) 257.}

\bref\Get{
E.~Getzler, {\it Batalin-Vilkovisky algebras and two-dimensional
topological field theories} ({\tt hep-th/9212043}). }

\bref\PS{
M.~Penkava and A.~Schwarz, {\it On some algebraic structures arising
in string theory} ({\tt hep-th/9212072}).}

\bref\Kr{
I.S.~Krasil'shchik, \LNM{1334} (1988) 79.}

\bref\Wia{
E.~Witten, \MPL{A5} (1990) 487.}

\bref\Th{
C.~Thielemans, \IJMP{C2} (1991) 787.}

%-----------------------------------------------------------------------------
%
%  TITLEPAGE
%

%\line{\hfil USC-94/17 }\bigskip
\line{\hfil } \bigskip

\leftline{{\bf THE BV-ALGEBRA STRUCTURE OF $\cW_3$ COHOMOLOGY}
\footnote{$^\flat$}{Published in the Proceedings of
``G\"ursey Memorial Conference I: Strings and Symmetries,''
{\it eds.}  G.~Akta\c{s} et al.,
Lect.\ Notes in Phys.\ {\bf 447} (Springer Verlag, Berlin, 1995).}}\vskip1cm

\leftline{Peter Bouwknegt and Krzysztof Pilch}
\vskip.3cm

\leftline{Department of Physics and Astronomy, U.S.C., Los Angeles, CA
90089-0484}\vskip1cm

\noindent {\bf Abstract:}
We summarize some recent results obtained in collaboration with J.~McCarthy
on the spectrum of physical states in $\cW_3$ gravity coupled to $c=2$ matter.
We show that the space of physical states, defined as a semi-infinite
(or BRST) cohomology of the $\cW_3$ algebra, carries the structure of
a BV-algebra. This BV-algebra has a quotient which is isomorphic to
the BV-algebra of polyvector fields on the base affine space of $SL(3,\CC)$.
Details will appear elsewhere.

%-----------------------------------------------------------------------------
%  1. INTRODUCTION
%
\newsec{Introduction}

Understanding the spectrum of physical states in theories
of two-dimensional $\cW$-gravity coupled to matter poses
an interesting challenge. Unlike in the case
of ordinary gravity, the computation of the relevant semi-infinite
(or BRST) cohomology of the underlying $\cW$-algebra appears to be very
difficult, and only a small number of results have been rigorously
established. One expects that by studying the
structure of this cohomology space it might be possible to achieve
a better understanding of (quantum) $\cW$-geometry and string field theory.
The problem is also mathematically quite interesting as it involves
generalizing some of the standard techniques for computing
semi-infinite cohomologies to non-linear
algebras.

In this paper we summarize some recent work done in collaboration
with J.~McCarthy on the computation of physical states in $\cW_3$-gravity
coupled to two scalar fields, as the semi-infinite cohomology of a
tensor product of two Fock space modules of the $\cW_3$ algebra. A complete
result for the cohomology is given in Conjecture 3.1, Theorem 3.2 and
Corollary 3.3. We then discuss in some detail the structure of the space
of physical states as a Batalin-Vilkovisky (BV) algebra and, in
particular, show that it is modelled on the well-known BV-algebra
of regular polyvector fields on the base affine space of $SL(3,\CC)$.
The main result here is given in Theorem 4.6. For more details
we refer to [\BMPa--\BMPc] and the forthcoming paper [\BMPd].

Throughout this paper we will use the notation $\bfh$ for
the Cartan subalgebra, $\bfh^*_\ZZ$ for the set of integral weights,
$P_+$ for the set of dominant integral weights,
$P_{++}$ for the set of strictly dominant integral weights,
$\De_+$ for the positive roots and $W$ for the Weyl group
of some Lie algebra $\bfg$. $\cL(\La)$ will denote the finite dimensional
irreducible representation of $\bfg$ with highest weight $\La\in P_+$ and
$\ell(w)$ the length of $w\in W$. In the following $\bfg$ will always
refer to $\slth$.

%-----------------------------------------------------------------------------
%  2. THE W_3 ALGEBRA AND ITS MODULES
%
\newsec{The $\cW_3$ algebra and its modules}

The $\Wt$ algebra with central charge $c\in\CC$
(denoted simply by $\cW$ in the sequel)
is defined as the quotient of the
free Lie algebra generated by  $L_m$, $W_m$,
$ m\in\ZZ$, by the ideal generated by
the following commutation relations (see \eg the
review on $\cW$-algebras [\BS], and references therein).
\eqn\eqAa{\eqalign{
[L_m,L_n]&=(m-n)L_{m+n}+\textstyle{c\over 12}
m(m^2-1)\de_{m+n,0}\,,\cr
[L_m,W_n]&= (2m-n) W_{m+n}  \,, \cr
[W_m,W_n]&=  (m-n)\left( \textstyle{1\over 15}
(m+n+3)(m+n+2)-\textstyle{1\over 6}
(m+2)(n+2)\right)L_{m+n} \cr
& +\be (m-n) \La_{m+n} +\textstyle{c\over 360}m(m^2-1)(m^2-4)\de_{m+n,0}
	     \,, \cr}
}
where $\be=16/(22+5c)$ and
\eqn\eqAb{
\La_{m}=\sum_{n\leq-2} L_{n} L_{m-n} +  \sum_{n>-2} L_{n} L_{m-n}
  -\textstyle{3\over 10}(m+3)(m+2) L_m\,.
}
Notice that, due to the non-linearity of $\La_m$ in \eqAa, $\cW$ is {\it not}
a Lie algebra.
The Cartan subalgebra $\cW_0$ of $\cW$ is spanned by
$ L_0$ and $ W_0 $, but, because $({\rm ad}\,W_0)$
is not diagonalizable,  $\cW$ does not admit a root space decomposition
(a generalized root space decomposition, \ie a Jordan normal form, does
however exist).
Nevertheless, it is still convenient to decompose the generators
of $\cW$ according to the  $(-{\rm ad}\,L_0)$ eigenvalue, and define
$\cW_{\pm}=\{L_n,W_n\,|\, \pm n>0\}$.
However,
this is not a triangular decomposition in the usual sense.

For physical applications the most interesting representations of $\cW$
are the so-called positive energy
modules, which  are defined by the condition that
(the energy operator) $L_0$ is diagonalizable
with finite dimensional eigenspaces, and with the spectrum bounded from below.
If the lowest energy eigenspace is one dimensional, we
denote the eigenvalues of $L_0$ and $W_0$
on the highest weight state by $h$ and $w$, respectively.

In particular, the Verma module $M(h,w,c)$ is defined
as the (positive energy) module induced by $\cW_{-}$ from an $1$-dimensional
representation of $\cW_0$.
By the standard argument, $M(h,w,c)$ contains a maximal
submodule. We denote the corresponding irreducible quotient
module by $L(h,w,c)$.
The module contragradient to $M(h,w,c)$ will be denoted by $\ffM(h,w,c)$.

Another class of positive energy
modules of $\cW$ are the Fock space modules $F(\La,\al_0)$, which
arise in the free field realization of $\cW$ in terms of two scalar fields
(see \eg [\BS], and references therein).
The modules $F(\La,\al_0)$ are labelled by the background charge
$\al_0\in\CC$ and an $\slth$ weight $\La$.

The central charge $c$ and
the highest weights $h$ and $w$ of $F(\La,\al_0)$ are given by
\eqn\eqAh{\eqalign{
c(\al_0) & = 2-24\al_0{}^2\,, \cr
h(\La)&=-(\th_1\th_2+\th_1\th_3+\th_2\th_3)-\al_0{}^2=\half(\La,\La+
2\al_0\rh)\,,\cr
w(\La)&=\sqrt{3\be}\,\th_1\th_2\th_3\,,\cr}
}
where
\eqn\eqAi{
\th_1=(\La+\al_0\rh,\La_1)\,,\quad
	    \th_2=(\La+\al_0\rh,\La_2-\La_1)\,,\quad
	    \th_3=(\La+\al_0\rh,-\La_2)\,.}
Here, $\La_1$ and $\La_2$ are the fundamental weights of $\slth$, and
$\rh=\half\sum_{\al\in\De_+}\al$ is the Weyl vector.
Note that $h(\La)$ and $w(\La)$ as in  \eqAh\  determine $\La$ only up to a
Weyl rotation $\La\rightarrow w(\La+\al_0\rh)-\al_0\rh$, $w\in W$.

The following theorem summarizes some of the known results on the structure
of Fock space modules $F(\La,\al_0)$:
\thm\thAa
\proclaim Theorem \thAa\ [\BMPa,\BMPb].
\item{(i)} Let $\imath'$ and $\imath''$ be the
canonical ($\cW$-) homomorphisms
\eqn\eqAc{ \matrix{
M(h(\La),w(\La),c(\al_0)) & \mapright{\imath'} & F(\La,\al_0) &
  \mapright{\imath''} & \ffM(h(\La),w(\La),c(\al_0)) \,. \cr}
}
Then $\imath'$ (resp.\ $\imath''$) is an isomorphism
if $i(\La+\al_0\rh) \in \et D_+$ (resp.\  $-i(\La+\al_0\rh) \in \et D_+$ )
and $\al_0{}^2 \leq-4$. Here $D_+ = \{ \la\in\bfh^* | (\la,\al)\geq0\
\forall \al\in\De_+\}$ denotes
the fundamental Weyl chamber and $\et \equiv {\rm sign}(-i\al_0)$.
\item{(ii)} For $c=2$,
the Fock space $F(\la,0)$ is completely reducible. Explicitly,
for all $\la\in \bfh^*_\ZZ$, we have
\eqn\eqAd{
F(\la,0)\ \cong\ \bigoplus_{\La\in P_+}\ m^\La_\la\, L(h(\La),w(\La),2)\,,
}
where $m^\La_\la$ is equal to the multiplicity
of the weight $\la$ in the irreducible
finite dimensional representation $\cL(\La)$
of $\slth$ with highest weight $\La$.\par

%-----------------------------------------------------------------------------
%  3. FOCK SPACE COHOMOLOGY OF THE W_3 ALGEBRA
%
\newsec{Fock space cohomology of the $\cW_3$ algebra}

Despite the fact that $\cW$ is not a Lie algebra, the analog of
semi-infinite (or BRST-) cohomology can still be defined [\TM,\BLNW].
As usual, one introduces
two sets of ghost operators $(\ghb{j}_m,\ghc{j}_m)\,,j=2,3$
of conformal dimension $(j,-j+1)$,
corresponding to the generators $L_m$ and $W_m$, $m\in \ZZ$, respectively.
These ghost operators satisfy
anti-commutation relations $\{ \ghb{j}_m, \ghc{j'}_n\} =
\de_{m+n,0}\de^{j,j'}$. Let $F^{\rm gh}$ denote the standard positive
energy module. The ghost Fock space $F^{\rm gh} = \bigoplus_{n\in\ZZ}
F^{{\rm gh},n}$ is graded by ghost number, where
${\rm gh}(\ghc{j}_m) = - {\rm gh}(\ghb{j}_m) = 1$ and the highest weight
state (physical vacuum) is chosen to have ghost number $3$ (\ie
such that states
and their corresponding operators have identical ghost numbers).
For any two positive energy modules $V^M$ and $V^L$, such that
$c^M + c^L = 100$, there exists a complex $(V^M \otimes V^L \otimes
F^{{\rm gh},n}, d)$, graded by ghost number, and with a differential
(BRST operator) $d$ of degree $1$. For an explicit formula for $d$,
which is rather involved, we refer to [\BLNW,\BMPa,\BMPb].
We will denote the cohomology
of this complex by $H(\cW, V^M \otimes V^L)$. The cohomology relative
to the Cartan subalgebra $\cW_0$ will be denoted by
$H(\cW,\cW_0; V^M \otimes V^L)$.

For $V^L \cong F(\La^L,\al_0^L)$ this cohomology
is interpretated as the
set of physical states in $\cW$-gravity coupled to some matter theory
represented by $V^M$. One is interested mainly in two cases:
where $V^M$ is either a so-called minimal model $L(h^M,w^M,c^M)$ or
a free field Fock space $F(\La^M,\al_0^M)$.
The minimal model case was discussed
in [\BMPa,\BMPc]. The analysis of
$H(\cW, F(\La^M,\al_0^M) \otimes F(\La^L, \al_0^L))$
for generic $\al_+$ (\ie $\al_+{}^2\notin\QQ$ where
we have parametrized $\al_0^M = \al_+ + \al_-\,,-i\al_0^L = \al_+ - \al_-\,,
\al_+\al_-=-1$) was started in [\BLNW] and completed in [\BMPc].
Here we will complete the analysis, begun in [\BMPb], of a non-generic
case, namely $\al_\pm = \pm1$ (\ie
$\al_0^M=0, -i\al_0^L =2$ or $c^M=2, c^L=98$).

Because of Theorem \thAa\ (ii) it suffices to compute the cohomology for
the $c=2$ irreducible $\cW$-modules $L(\La) \equiv L(h(\La),w(\La),2)$
\thm\thBa
\proclaim Conjecture \thBa\ [\BMPd].
Let $\La\in P_{+}$.
\item{(i)} The cohomology $H^n(\cW,\cW_0;L(\La) \otimes F(\La^L,2i))$ is
nontrivial only if there exist $w \in W,\ \si\in W\cup \{0\}$ such that
\eqn\eqBa{
-i\La^L + 2 \rh = w^{-1} (\La + \rh - \si \rh)\,.
}
\item{(ii)}
For $w,\si,\La$ and $\La^L$ as in \eqBa,
the cohomology $H^n(\cW,\cW_0;L(\La) \otimes F(\La^L,2i))$
is 1-dimensional in the following cases
$$\matrix{
\si \in W\,,  &  \La\in P_+\,,&  w\in W\,, &
n= \ell(w^{-1}) - \ell(w^{-1}\si) +3\,, \cr
\si =0\,,     &   \La\in P_{++}\,, &  w\in W\,,   &
n=\ell(w^{-1})+1 {\rm \ or\ } n=\ell(w^{-1})+2 \,, \cr
\si=0\,,    &  (\La,\al_i) = 0\,,\La\neq0\,,  &
w\in\, <\!r_i\!>\!\backslash W\,, &   n=\ell(w^{-1})+2\,,\cr
\si=0\,, & (\La,\al_i) = 0\,,\La\neq0\,, &
w\in r_i(<\!r_i\!>\!\backslash W)\,,  & n=\ell(w^{-1})+1\,. \cr}
$$
and vanishes otherwise.

\noindent {\sl In the case that certain weights $(\La,-i\La^L)$ and
certain ghost
number $n$ satisfy {\sl (i)} and {\sl (ii)} for more than
one choice of $(w, \si)$, the above should be understood in the
sense that the
corresponding cohomology is nevertheless 1-dimensional.}

Let us comment on the status of this conjecture. For
$-i\La^L + 2 \rh \in P_+$ we have
an isomorphism $F(\La^L,2i) \cong \ffM(h(\La^L),w(\La^L),2)$
(see Theorem \thAa\ {\sl (i)}). By taking the (conjectured) resolutions
of $L(\La)$ in terms of generalized Verma modules $M(h,w,c=2)_N$ [\BMPb]
and using the known result for
$H^n(\cW,\cW_0; M(h,w,c) \otimes \ffM(h',w',100-c))$, the conjecture
follows (see [\BMPb] for details).
[The resolution of $L(\La)$ for $\La\in P_{++}$ in [\BMPb]
contains a minor misprint, see [\BMPd].]

For the other Weyl chambers, \ie $w(-i\La^L + 2\rh) \in P_+$, the
conjecture is based on an analysis of the cohomology for generic $\al_+$
in the limit $\al_+ \to1$ (\ie $c^M\to2$) and passes various nontrivial
consistency checks. Among others, it is consistent with duality
\eqn\eqPc{
H^{6-n}(\cW,\cW_0;L(\La) \otimes F(\La^L,2i)) \cong
H^{n}(\cW,\cW_0;L(\La) \otimes \ffF(\La^L,2i))\,,
}
where $\ffF(\La,\al_0) \cong F(w_0(\La + \al_0\rh) - \al_0\rh)$ denotes
the module contragradient to $F(\La,\al_0)$.

Both the conjectured resolutions of $L(\La)$ as well as the result
for the cohomology (Conjecture \thBa) have also been verified by
extensive computer calculations using Mathematica$^{\rm TM}$.

Let $L$ be the lattice
\eqn\eqBo{
L\ \equiv\ \{ (\la,\mu) \in \bfh^*_\ZZ \otimes \bfh^*_\ZZ\,|\,
  \la - \mu \in \ZZ\cdot\De_+ \} \,.
}
Note that, in particular,
\eqn\eqBq{
(\la,\la') - (\mu,\mu') = (\la-\mu,\la') + (\mu,\la'-\mu') \in \ZZ\,,
}
for all pairs $(\la,\mu)$ and $(\la',\mu')$ in  $L$.
We will restrict the momenta $(\La^M,-i\La^L)$ to the lattice $L$.
As a consequence, all the vertex operators
$V_{(\La^M,\La^L)}(z) = \exp(i\La^M\cdot \ph^M + i\La^L\cdot\ph^L)(z)$
will become mutually local because of \eqBq\ and, moreover, one
can find a set of cocycles turning the underlying BRST-complex
into a Vertex Operator Algebra (VOA). This will be essential for the
construction of the BV-algebra in Section 3. In addition,
the most interesting cohomology happens to be situated at
$(\La^M,-i\La^L)\in L$.

Now consider the cohomologies
\eqn\eqBp{ \eqalign{
\cH  & = \bigoplus_{ (\La^M,-i\La^L) \in L} H(\cW, F(\La^M,0) \otimes
  F(\La^L,2i))\,, \cr
\cH_{\rm rel}  & = \bigoplus_{ (\La^M,-i\La^L) \in L} H(\cW,\cW_0;
  F(\La^M,0) \otimes F(\La^L,2i))\,. \cr}
}
We recall
\thm\thBb
\proclaim Theorem \thBb\ [\BMPa,\BMPb].
\item{(i)} $\cH$ (and $\cH_{\rm rel}$) carries the structure of a
$\bfg\oplus\bfh$ module ($\bfg\cong\slth$). The action of $\bfg$ is
through the zero modes of the Frenkel-Kac-Segal vertex operator construction
(in matter fields only), while $\bfh$ acts as $-ip^L$ (with eigenvalues
$-i\La^L$). This module is completely reducible under $\bfg\oplus\bfh$.
\item{(ii)} There exists a (non-canonical) isomorphism (as $\bfg\oplus
\bfh$ modules)
$$
\cH^i\ \cong\ \cH_{\rm rel}^i \oplus \cH_{\rm rel}^{i-1}
  \oplus \cH_{\rm rel}^{i-1} \oplus \cH_{\rm rel}^{i-2} \,.
$$\par

By combining the results of Theorems \thAa, \thBb\ and Conjecture \thBa,
we find
\thm\thBc
\proclaim Corollary \thBc. The cohomology $\cH_{\rm rel}$ is isomorphic
(as a $\bfg\oplus \bfh$ module) to the direct sum of irreducible
modules $\cL(\La)\otimes \CC_{\La'}$ with momenta
$(\La,\La')\in \bfh^*_\ZZ \otimes \bfh^*_\ZZ$ lying in
a set of disjoint cones $\{ \cS^n_w + (\la,w^{-1}\la)\,|\,\la\in P_+\}$, \ie
$$
\cH_{\rm rel}^n\ \cong\ \bigoplus_{w\in W} \bigoplus_{(\La,\La')\in
  \cS_w^n} \bigoplus_{\la\in P_+} \left( \cL(\La+\la) \otimes
  \CC_{\La' + w^{-1}\la} \right) \,,
$$
where the sets $\cS_w^n$ (tips of the cones) are given in Table 1.\par
\vfil\eject

%-------------------------------------------------------------------------
% isttable.tex   940922-1

\begintable
\quad $n$ \quad | \quad $w$ \quad | \quad $\cS_w^n$ \quad \crthick
$0$ | $1$ |  $(0,0)$  \cr
$1$ | $1$ |  $(\La_1,-\La_1+\La_2)$, $(\La_1+\La_2,0)$,
             $(\La_2,\La_1-\La_2)$  \nr
    | $r_1$ | $(0,-2\La_1+\La_2)$ \nr
    | $r_2$ | $(0,\La_1-2\La_2)$ \cr
$2$ | $1$ | $(2\La_1,-\La_1)$, $(0,-\La_1-\La_2)$,
            $(2\La_2,-\La_2)$ \nr
    | $r_1$ | $(\La_1,-2\La_1)$, $(\La_2,-3\La_1+\La_2)$,
              $(0,-4\La_1+2\La_2)$ \nr
    | $r_2$ | $(\La_2,-2\La_2)$, $(\La_1,\La_1-3\La_2)$,
              $(0,2\La_1-4\La_2)$ \nr
    | $r_2r_1$ | $(0,-3\La_1)$ \nr
    | $r_1r_2$ | $(0,-3\La_2)$ \cr
$3$ | $1$ | $(\La_1+\La_2,-\La_1-\La_2)$ \nr
    | $r_1$ |  $(\La_2,-2\La_1-\La_2)$, $(\La_1,-4\La_1+\La_2)$,
              $(\La_2,-5\La_1+2\La_2)$  \nr
    | $r_2$ | $(\La_1,-\La_1-2\La_2)$, $(\La_2,\La_1-4\La_2)$,
              $(\La_1,2\La_1-5\La_2)$ \nr
    | $r_2r_1$ | $(\La_1,-3\La_1-\La_2)$, $(0,-5\La_1+\La_2)$,
                 $(\La_1,-5\La_1)$ \nr
    | $r_1r_2$ | $(\La_2,-\La_1-3\La_2)$, $(0,\La_1-5\La_2)$,
                 $(\La_2,-5\La_2)$ \nr
    | $r_1r_2r_1$ | $(0,-2\La_1-2\La_2)$ \cr
$4$ |  $r_1$ | $(0,-4\La_1-\La_2)$ \nr
    | $r_2$ | $(0,-\La_1-4\La_2)$ \nr
    | $r_2r_1$ | $(\La_1,-4\La_1-2\La_2)$, $(\La_2,-5\La_1-\La_2)$,
                 $(0,-6\La_1)$ \nr
    | $r_1r_2$ | $(\La_2,-2\La_1-4\La_2)$, $(\La_1,-\La_1-5\La_2)$,
                 $(0,-6\La_2)$ \nr
    | $r_1r_2r_1$ |  $(0,-3\La_1-3\La_2)$, $(2\La_1,-4\La_1-3\La_2)$,
                $(2\La_2,-3\La_1-4\La_2)$ \cr
$5$ | $r_2r_1$ | $(0,-5\La_1-2\La_2)$ \nr
    | $r_1r_2$ | $(0,-2\La_1-5\La_2)$ \nr
    | $r_1r_2r_1$ |  $(\La_1,-5\La_1-3\La_2)$, $(\La_1+\La_2,-4\La_1-4\La_2)$,
        $(\La_2,-3\La_1-5\La_2)$ \cr
$6$ | $r_1r_2r_1$ | $(0,-4\La_1-4\La_2)$ \endtable
\medskip

\centerline{\it Table~1.\ The sets $\cS_w^n$}\medskip
%-------------------------------------------------------------------------

In particular we see that, as an $\slth$ module, the `ground
ring' $\cH^0$ decomposes as $\cH^0 \cong \bigoplus_{\La\in P_+}
\cL(\La)$ and is
therefore a so-called `model space' for $\slth$. It is well-known
that this model space can be realized as the space $\cP^0(A)$ of polynomial
functions on the so-called `base-affine space' $A\equiv N_+\backslash G$
[\BGG].
For $\slth$ this model space is given by $\CC\,[x^i,y_i] / \vev{x^iy_i}\
(i=1,2,3)$, \ie polynomials in $6$ variables $x^i, y_i$ transforming in
the $\bf 3$ and $\bar{\bf 3}$ of $\slth$ respectively, with a single
relation $x^iy_i=0$ [\Zh]. In fact, one can show that $\cH^0 \cong \cP^0(A)$
as algebras [\BMPd].
One might think that, just as in the Virasoro
case (corresponding to $\bfg\cong\sltw$) [\Wi--\LZ], part of the
rest of $\cH$ allow an interpretation in terms of polyvector fields
on this base affine space. This turns out to be true and will be elaborated
on in the next section.
\vfil\eject

%-----------------------------------------------------------------------------
%  4. THE BV-STRUCTURE OF H
%
\newsec{The BV-structure of $\cH$}

To explain the algebraic structure of the cohomology $\cH$ of Section 2
we will first need to recall the definition of a Gerstenhaber algebra
(or G-algebra, for short) [\Ge]
and a BV-algebra (or coboundary G-algebra)
[\Ko--\PS,\LZ] as well as some basic facts.
\thm\thCa
\proclaim Definition \thCa. A G-algebra $(\cA,\,\cdot\,,[\ \,,\ ])$ is a
$\ZZ$-graded, supercommutative, associative algebra
$\cA = \bigoplus_{i\in\ZZ} \cA^i$ (under $\cdot$) as well as a $\ZZ$-graded
Lie superalgebra (under $[\ \,,\ ]$), such that the (odd) bracket acts as a
superderivation of the algebra, \ie
\eqn\eqCa{
[x,y\cdot z] = [x,y]\cdot z + (-1)^{(|x|-1)|y|} y\cdot [x,z]\,,\qquad
x,y,z \in \cA\,.
}
\par

For any commutative algebra $\cA$ and $\cA$-module $\cM$, one defines
the the set $\cD(\cA,\cM)$ of derivations of $\cA$ with coefficients in
$\cM$ as the set of elements $D\in {\rm Hom}(\cA,\cM)$ that satisfy the
Leibniz rule
\eqn\eqCb{
D(x\cdot y) = y (Dx) + x (Dy)\,.
}
The set $\cD^n(\cA)$ of polyderivations of order $n$
is defined by induction as those $D\in {\rm Hom}(\cA,\cD^{n-1}(\cA))$
satisfying the Leibniz rule \eqCb\ as well as being completely antisymmetric
when considered as elements of ${\rm Hom}(\cA^{\otimes n},\cA)$. We recall
\thm\thCb
\proclaim Theorem \thCb\ [\Kr]. Let $\cA$ be a commutative
algebra. The set of polyderivations $\cD(\cA)$ carries the structure
of a G-algebra, with the bracket given by the Schouten bracket.\par

Another example of a G-algebra is the Hochschild cohomology $H(\cA,\cA)$
of an associative algebra $\cA$ [\Ge].
\thm\thCc
\proclaim Definition \thCc. A BV-algebra $(\cA,\,\cdot\,,\De)$
is a $\ZZ$-graded,
supercommutative, associative algebra $\cA$ with a second order
derivation $\De$ (BV-operator) of degree $-1$ satisfying $\De^2 = 0$.\par

\thm\thCd
\proclaim Lemma \thCd\ [\Wia,\LZ,\PS]. For any BV-algebra $(\cA,\,\cdot\,,\De)$
we may define an odd bracket by
\eqn\eqCc{
[x,y] = (-1)^{|x|} \left( \De(x\cdot y) - (\De x)\cdot y -
  (-1)^{|x|} x\cdot(\De y) \right)\,,\qquad x,y\in \cA\,.
}
This will equip $\cA$ with the structure of a G-algebra. Moreover, the
BV-operator acts as a superderivation of the bracket
\eqn\eqCd{
\De [x,y] = [\De x,y] + (-1)^{|x|-1} [x,\De y]\,.
}
\par

In general, given a commutative algebra $\cA$, the G-algebra $\cD(\cA)$
of polyderivations of $\cA$ will not carry the structure of a BV-algebra.
However, if $\cA$ is the algebra of (smooth or polynomial) functions on some
smooth manifold $M$, then $\cD(\cA)$ is isomorphic to the set of
polyvector fields $\cP(M)$ on $M$ [\Kr]. If, moreover,
$M$ possesses a volume form, then we can in fact equip $\cD(\cA)$ ($=\cP(M)$)
with the structure of a BV-algebra [\Wia,\LZ]. Another example of a BV-algebra
is the Grassmann algebra $\bigwedge^*\bfg$ of a Lie algebra $\bfg$ [\LZ].

Given a BV-algebra $(\cA,\,\cdot\,,\De)$, let $\cA^0$ be its `ground ring.'
It follows from equations \eqCa, \eqCc\ and \eqCd\ that
there exists a natural way to embed $\cA$ into the G-algebra of
polyderivations of $\cA^0$, \ie $\cD(\cA^0)$, namely
\thm\thCe
\proclaim Theorem \thCe. Let $(\cA,\,\cdot\,,\De)$ be a BV-algebra. Suppose
$\cA^n = 0$ for all $n<0$.
\item{(i)}There exists a homomorphism of G-algebras
$\pi:\cA \to \cD(\cA^0)$ defined by
\eqn\eqCe{
\pi(y) (x_1,x_2,\ldots,x_n) = [[\ldots [[y,x_1],x_2],\ldots],x_n]\,,\qquad
y\in \cA^n,\, x_1,x_2,\ldots,x_n\in \cA^0\,.
}
\item{(ii)} Suppose that the G-algebra $\cD(\cA^0)$ admits a BV-structure
$(\cD(\cA^0),\,\cdot\,,\De')$ and that $\pi\De(x) = \De'\pi(x)$ for
all $x\in \cA^1$, then $\pi$ is a BV-homomorphism and $\cI \equiv
{\rm Ker\,}\pi$ is a BV-ideal of $\cA$.\par

We are now ready to state the main result of this paper
\thm\thCf
\proclaim THEOREM \thCf. Let $\cH$ be the cohomology defined in \eqBp. Then
\item{(i)} $\cH$ can be equipped with the structure of a BV-algebra.
\item{(ii)} There exists an ideal $\cI \subset \cH$ such that
we have an exact sequence of BV-algebras
\eqn\eqCf{ \matrix{
0 & \mapright{} & \cI & \mapright{} & \cH & \mapright{\pi} &
\cD(\cH^0) & \mapright{} & 0\,,\cr}
}
where $\cD(\cH^0)$ is isomorphic
to the BV-algebra $\cP(A)$ of polyvector fields on the
base affine space $A = N_+\backslash G$.\par

Let us make some comments on the proof.
Quite generally, as has been shown in [\Wi--\LZ,\PS],
BRST cohomologies of VOA's carry the structure of a BV-algebra.
The product in this BV-algebra is given by the normal ordered product of
the VOA while $\De = b_0^{[2]}$.
The crucial part of the proof of {\sl (i)} is therefore to show that the
complex carries the structure of a VOA.
This amounts to showing that one can
find an appropriate set of cocycles for the lattice $L$. This is a
straightforward exercise.
[One might wonder whether there exists additional structure in $\cH$
beyond that of a BV-algebra, in particular whether
$\ghb{3}_0$ gives rise to a second BV-operator. It turns out however
that, due to the non-diagonalizability of $W_0$, $\ghb{3}_0$ does {\it not}
act on $\cH$.]
As we have
seen in Section 2, there exists a canonical isomorphism of algebras
$\cH^0 \cong \cP^0(A)$, where $\cP^0(A)$ denotes the
(commutative) algebra of polynomials on $A$.
This implies $\cD(\cH^0)\cong\cP(A)$ as algebras. That $\pi$ is in fact a
BV-epimorphism follows from Theorem \thCe\
by explicitly checking that $\pi$ intertwines
the BV-operators on $\cH^1$ and $\cP^1(A)$ and that it acts onto.

We would like to remark here that, contrary to the Virasoro case
[\LZ], both the dot product
and the bracket in $\cI$ are not identically zero.
Also, the exact sequence \eqCf\ splits both as an
exact sequence of $\cH^0$ and $\bfg\oplus
\bfh$ modules, but {\it not} as an exact sequence of BV-algebras.

Details of this paper as well as a more detailed analysis of the BV-algebra
structure of the entire $\cH$ will appear elsewhere [\BMPd].
\bigskip

%-----------------------------------------------------------------------------
%
%  ACKNOWLEDGEMENT
%
\noindent{\bf Acknowledgement:} We would like to thank the organizers
of the ``1st G\"ursey Memorial Conference'' for the opportunity to
present this talk, and C.~Thielemans for making
available to us his Mathematica package {\tt OPEdefs} [\Th].

%-----------------------------------------------------------------------------
\footatend\immediate\closeout\rfile\writestoppt
\baselineskip=14pt{\bigskip\noindent {\bf  References}}%
\bigskip{\frenchspacing%
\parindent=20pt\escapechar=` \input refs.tmp\vfill\eject}\nonfrenchspacing
\vfil\eject\end